\documentclass[11pt]{article}
\usepackage{setspace}
\usepackage{amssymb}
\usepackage{amsmath}
\usepackage{amsfonts}
\usepackage{slashed}

\def\be{\begin{equation}}

\def\ee{\end{equation}}

\def\dd{\partial}

\def\bea{\begin{eqnarray}}

\def\eea{\end{eqnarray}}

\newcommand\eps{\epsilon}

\makeatletter
\def\blfootnote{\xdef\@thefnmark{}\@footnotetext}
\makeatother

\begin{document}

\singlespace

\begin{flushright} BRX TH-6650 \\
CALT-TH 2019-019
\end{flushright}

\vspace*{.3in}

\begin{center}

{\Large\bf Noether Converse: all identically conserved geometric tensors are metric variations of an action in (at least) $D=2$}

{\large S.\ Deser}

{\it 
Walter Burke Institute for Theoretical Physics, \\
California Institute of Technology, Pasadena, CA 91125; \\
Physics Department,  Brandeis University, Waltham, MA 02454 \\
{\tt deser@brandeis.edu}
}
\end{center}

\begin{abstract}
The title's century-old conjecture is established for $D=2$ and is likely for all $D$.
\end{abstract}

\section{Introduction}

The physical impact of Noether's theorems is undimmed after a century, yet the second's (local symmetries) equally old converse has remained unsolved, despite countless arduous attempts. [For a very recent discussion, technical details and historical references, see [1], where the most advanced previous result is proved: the converse is valid to order $\dd^6 g_{\mu\nu}$.] The theorem of course states that the variation of every locally gauge invariant action is identically conserved; more colloquially, that the gauge variation of a gauge invariant vanishes, and conversely if it vanishes, its action is invariant. For vector gauge invariant actions such as Maxwell, this is obvious by antisymmetry of the field strengths since the variation of the potential $A_\mu$ of any action $\int L(F^{\mu\nu})$ is a vector $J^\mu$ proportional to $\dd_\nu H^{[\mu\nu]}$, where $H$ is necessarily antisymmetric, being an $F$-variation. The nonabelian, YM, case is identical, despite appearance of covariant color derivatives, because of the structure constants' antisymmetries. However, it is known that the converse fails here, ironically for the same reason: there are infinitely many identically conserved $J^\mu$ that are NOT action-variations, e.g., $J^\mu = \dd_\nu [(F \,^*F)F^{\mu\nu}]$, where $\, ^* F$ is the $D=4$ dual of $F$. In other words, any ``superpotential" $\dd_\nu H^{[\mu\nu]}$ is trivially annihilated by $\dd_\mu$ (or even by $D_\mu$ for YM), independent of $H$'s non/action origin.

The geometric, coordinate invariant, case is another story, however, because covariant derivatives now do matter. For the abelian limit, things are still as for vectors because there is still an identically conserved symmetric superpotential, $Z^{(\mu\nu)}= \dd^2_{\alpha\beta} H^{[\mu \alpha][\nu \beta]}$, where $H$ shares the algebraic symmetries of the Riemann tensor. So any abelian tensor current $Z^{(\mu\nu)}$ is again manifestly identically (ordinarily) conserved for any, also non-action, $H$. However, the similarity ceases for full general covariance, where the now covariant divergence is no longer identically conserved: the commutator of two covariant derivatives is proportional to the curvature. Does this imply that the converse holds here? That is the question. To avoid index proliferation, I work explicitly in $D=2$, where only the scalar curvature enters, then indicate why the result should hold for all $D$.

\section{Gravities}
In the abelian limit and in $D=2$, $H$ degenerates to $\eps^{\mu\alpha} \eps^{\nu\beta} \dd^2_{\alpha\beta} S$, namely to the transverse projector, for arbitrary scalar $S$,
\begin{equation}
O^L_{\mu\nu} S = (\dd^2_{\mu\nu} -\eta_{\mu\nu} \Box) S.
\end{equation}
The linear $O^L$ is proportional to $\delta R^L/\delta h_{\mu\nu}$, and the full $O^{\mu\nu}$ is correspondingly proportional to $\delta R/\delta g_{\mu\nu}$ where all derivatives in (1) are covariant, so its divergence now becomes 
\begin{equation}
D_\nu O_\mu^\nu S= [D_\nu D_\mu - D_\mu D_\nu] \dd^\nu S = [D_\nu, D_\mu]\dd^\nu S =R \dd_\mu S \ne 0 \rightarrow S \dd_\mu R\ne 0;  
\end{equation}
the last inequality follows from adding a $g_{\mu\nu} RS$ term, as would arise from varying the curvature density rather than the scalar in $O$. The divergence of this $\delta A/\delta R \delta R/ \delta g_{\mu\nu}$ part of an action's variation is canceled by that of its explicit metric variation $\delta A/\delta g_{\mu\nu}$, in covariant derivatives and contracting metrics; that's just Noether. But might there be a compensating tensor $X^{\mu\nu}$ whose divergence $D^\nu X_{\mu\nu}$ is $-S \dd_\mu R$ for some S, absent an action?  Or indeed, might there be a conserved tensor that is entirely different, independent of the existence of any projector? We will first eliminate any that contains a term $g^{\mu\nu} Q$, the hallmark of any action-based solution, since these always\footnote{Parity violating terms, linear in the Levi-Civita density $\eps^{\alpha\beta\ldots}$, do not have a $\sqrt{-g}$ in their action, but our $D=2$ $X^{\mu\nu}$ can only maintain symmetry if $\eps^{\alpha\beta}$ is absent: any ``internal" $\eps$ in one $D^k R$ chain forces it in the other, losing the density weight. In $D=3$, there are Chern-Simons (action) gravities, for example; we have not studied if there are dangerous candidate non-action $X^{\mu\nu}$ in $D>2$, but it seems unlikely. } contain a $\sqrt{-g}$ term, $A= \int \sqrt{-g} Q$. We will then exclude candidates containing the (ex-)projectors $O$, leaving finally the ``hard core" pure $Y$ case.

The general $X^{\mu\nu}$ may be written as $1/2 g^{\mu\nu} Q +Y^{\mu\nu}$, where $Y$ is a symmetric tensor, both of whose open indices must therefore be derivatives on some $D^k R$, and $Q$ is a (possibly $0$) scalar. [The $g_{\mu\nu} Q$ term is unambiguous --- it is not equivalent to some other tensor by algebraic identities.]  The $X=g Q$ part is the metric variation of $\sqrt{-g}$ in the action  $\int d^2 x \sqrt{-g} Q$, so $(X+Y)$ is action-generated IFF $Y$ is $Q$'s total metric variation, i.e., through its $D^k$ as well as its explicit metric-dependence. [Currents independent of curvature are easy: there is only one in any $D$, namely the metric, whose action, is of course the volume integral.] If $Y$ is not the explicit metric variation of $ \int Q \sqrt{-g}$, we merely add and subtract the required new $Y'$ to restore the overall action status, tuning $Y'$ to the chosen $Q$, which is always possible, since $\int Q \sqrt{-g}$ is an allowed action for all scalar $Q$. Then $(X+Y')$ reduces to the above action case, leaving the difference between the original and the new $Y$.  So the question is now whether any purely $Y$-type tensor can be identically conserved. We next show that such $Y$ cannot contain any $OS$ contribution either. Absent the $g^{\mu\nu} Q$ term, $Y$'s open indices can only appear as derivatives on various $D^k R$. But $Y'$s $n$-divergence must have the required $-S \dd_\mu R$ form to compensate for that of the putative $OS$. Hence it must include the term $\sim  A(D^k R; g) R_{,\mu} R_{,\nu}$ since taking the divergence cannot lower a tensor's derivative rank, and $Y$ is symmetric. The $n$-divergence of $A R_{,\mu} R_{,\nu}$ does give the desired $\dd_\mu R$ dependence--- but also the unwanted term $A R_{,\nu} R_{,\mu\nu} = 1/2 A (R_{,\nu})^2_{,\mu}$; that one can only be removed by adding the, already excluded, $Q$-form $-1/2 g_{\mu\nu} A (R_{,\rho})^2$. But this in turn adds $A_{,\mu} (R_{,\rho})^2$, requiring $A=A(R)$, so again there's an action, $ \int d^2x (f_{,\nu})^2$, $f\sim  \int \sqrt{A} dR$.  Adding instead a term $\sim B R_{,\mu\nu}$ cannot help, because for example its divergence would include the non-cancelable third derivative $B R_{,\mu\nu}^{\, \, \, \, \, \, \nu}$. This underlines the difference between the case where $g Q$ is present, so the --- essential --- divergence $D_\mu$ acts on $Q$, and that where it is absent, and there remains a useless $D_\nu$. Now that we have excluded $OS$ as well, there remain the ``pure`` $Y$ tensors, those that contain neither $g Q$ nor $OS$. The  absence of $OS$ currents will be critical in this final stage. The most general symmetric $Y_{\mu\nu}$ (except for simplifying slightly by setting the derivative count equal, i.e., both $A$ coefficients have $k$ implicit covariant derivatives denoted as $R^k$; the thereby excluded sector cannot work either) has the form
\begin{equation}
Y_{\mu\nu} = A D_\mu R^\kappa D_\nu R^\kappa + B (D^2_{\mu\nu} +D^2_{\nu\mu})R^l.   
\end{equation}
In this highly condensed notation, $A/B$ are tensors contracting with all the implicit derivative indices on their $R^\kappa/R^\iota$ companions; we have moved the open indices uniformly to the end of each derivative string on the $R$, as best candidates; any other position only makes things worse, as it requires index interchanges, so invokes an immediate $[D,D]$. Can $D_\nu Y^{\mu\nu}$ be made to vanish for some $(A,B)$ and $(\kappa, \iota)$ derivative powers? It reads
\begin{equation}
D_\nu Y^\nu_{\mu}= D^\nu (A D_\nu R^\kappa)D_\mu R^\kappa +A D^\nu R^\kappa D^2_{\nu\mu} R^\kappa + D_\nu B (D^2_{\mu\nu} + D^2_{\nu\mu})R^\iota + B(D^\nu_{\, \, \mu\nu}+ D^\nu_{\, \, \nu\mu})R^\iota. 
\end{equation}
The problem is already clear here: we saw that no conserved $Y$ can have an $OS$ term, one formed by moving covariant derivatives at the price of a $[D,D]$ commutator (except when they act on a pure $D^0 R$, in which case (4) is easily seen to be nonzero), yet there are necessarily such parts of (4) --in particular the last, $B D^\nu_{\nu\mu} R^\iota$ term, where the open index ($\mu$) is ``hidden" behind two outer $D^2_\nu$, so it has no $A$-counterpart. That forces $B$ to vanish, leaving the $A$ part, whose two terms manifestly cannot cancel each other, given the open index's different positions, that would require (at least) a derivative commutation. That completes the $D=2$ story.\footnote{ A simple, but $D=2$ specific, proof uses conformal gauge, $g_{\mu\nu}= e^{2 \phi} \eta_{\mu\nu}$ to convert the covariant divergence identity to an ordinary derivative one, $\dd_\nu T^{\mu\nu} = \dd^\mu \phi Z$ that is almost manifestly unfulfillable by any $T$. Separately, note that while we started from the ``$X$-side" of the problem, rather than the "$S\dd R$ side", there is no loss of generality, because we were seeking ANY $S$ that would do the trick, i.e., any candidate from the $S\dd R$ side. 
}

\section{Higher $D$}
For $D>2$, the elimination of $g_{\mu\nu} Q$ is manifestly unchanged, while that of (ex-)projector generalizations of $OS$, namely $DDHS$, should also go through: even though they have more indices, the process is the same. We have not checked this in detail because there is a worse complication: open indices can now reside on curvature/Ricci tensors (though some can be turned into derivatives by cyclic identities). Absent the tedious explicit process of checking all index proliferations in any $D$ (or a better method!), a proof cannot yet be claimed for general $D$, but the analogies are pretty persuasive, at least to the author.

\section{Coda}
Our result saves trees (certainly for $D=2$, and probably for all $D$) by eliminating any would-be gravitational industry with non-action field equations, since in presence of normal sources from invariant actions (if those cannot be included, there is no physics) --- the matter side remains conserved on its shell and so therefore must the geometric one.

\subsubsection*{Acknowledgements}
This work was supported by the U.S. Department of Energy, Office of Science, Office of High Energy Physics, under Award Number 
de-sc0011632. I thank Y. Pang for many informative discussions and for saving me from egregious errors (if not from present ones) on this long journey; A. Waldron's input was also of great help, as was J. Franklin's tex skill.

\end{document}